\documentclass[fleqn,10pt]{wlscirep}
\usepackage[utf8]{inputenc}
\usepackage[T1]{fontenc}
\usepackage{float}
\usepackage{amsmath}
\usepackage{bigstrut}
\usepackage{commath}
\usepackage{csquotes}
\usepackage{ragged2e}
\usepackage{multirow}
\usepackage[font=footnotesize]{caption}
\usepackage{caption}
\usepackage{tabularx}
\usepackage[numbers,sort&compress]{natbib}
\usepackage{booktabs,makecell}

\title{Correcting temporal bias in mobility data using time-use surveys}

\author[1,2,3*]{Sarah A. Sanchez}
\author[3,4]{Hamish Gibbs}
\author[5]{Takahiro Yabe}
\author[1,2]{Daniel T. O'Brien}
\author[3,4]{Esteban Moro}

\affil[1]{School of Public Policy and Urban Affairs, Northeastern University, Boston, Massachusetts, USA}
\affil[2]{Boston Area Research Initiative, Boston, Massachusetts, USA}
\affil[3]{Social Urban Networks Lab, Boston, Massachusetts, USA}
\affil[4]{Network Science Institute, Northeastern University, Boston, Massachusetts, USA}
\affil[5]{Tandon School of Engineering, New York University, New York, USA}

\affil[*]{sa.sanchez@northeastern.edu}

\begin{abstract}

GPS mobility data is a valuable source of behavioral measurement which is subject to systematic biases including the over- or under-representation of demographic groups, and variations in the quality of location sampling across time. In this paper, we address the challenge of temporal bias in mobility data, which can skew the representation of mobility behaviors due to the event-based nature of location data sampling. We use the American Time Use Survey (ATUS) to assess the accuracy of a place-based measure of economic segregation drawn from large-scale mobility data across 11 U.S. cities. We show that comparisons with high quality time use surveys such as the ATUS can validate behavioral insights from mobility data, while quantifying uncertainty and highlighting areas of relative instability in analytical findings. We also propose a temporal re-weighting method that can complement existing bias-mitigation techniques to improve the accuracy of conclusions drawn from GPS-based mobility data.

\end{abstract}

\begin{document}

\maketitle

\section*{Introduction}
GPS mobility data from mobile phones have transformed our ability to observe micro-scale movement and activities in cities across millions of devices. As such, urban scientists have used them to illuminate various aspects of human behavior and experiences, including racial and socioeconomic segregation \cite{moro_mobility_2021, yabe_behavioral_2023, athey_estimating_2021, xu_using_2025}, public health \cite{tizzoni_use_2014, wesolowski_connecting_2016, buckee_aggregated_2020}, impacts of natural disasters \cite{yabe_mobile_2022, podesta_quantifying_2021, coleman_lifestyle_2023}, economic development \cite{wang_infrequent_2024, yabe_behaviour-based_2024}, and more. Though mobility data offer unique opportunities, they have well-documented limitations that can reduce their accuracy \cite{yabe_enhancing_2024, lazer_computational_2020}. While some of these limitations are technical in nature, such as measurement errors in GPS accuracy or systematic gaps in data collection, the more consequential challenges concern the extent to which GPS traces suffer from bias in the populations (i.e., \textit{demographic bias}) and behaviors (i.e., \textit{temporal bias}) that they capture (see Table \ref{tab:bias_grouped}).

Current mobility data research often acknowledges and investigates demographic bias, where mobility data may over- or under-represent specific demographic groups \cite{li_understanding_2024, moro_mobility_2021,cabrera_systematic_2025}. The importance of doing so was reinforced during the COVID-19 pandemic, when mobility measures failed to capture the activity of low-income workers, older adults, and non-white populations, groups disproportionately impacted by COVID-19 and associated lockdown measures \cite{kissler_reductions_2020, whittle_ecological_2020,coston_leveraging_2021, Jardel2024Uncovering, safegraph_what_2019, li_understanding_2024}. Similar limitations affect the representation of low-income and rural populations in mobility datasets globally \cite{grantz_use_2020}. Researchers and companies have increasingly adopted a range of methods to mitigate it \cite{salganik_bit_2017}, including post-stratification \cite{moro_mobility_2021}, pre-stratification \cite{klein_characterizing_2024}, synthetic data-augmentation \cite{berke_generating_2022,aleta_modelling_2020,aleta_quantifying_2022}, and validation against independent datasets or ground truth  \cite{pappalardo_evaluation_2021, eagleston_features_2024, moro_mobility_2021}. Among these, post-stratification is the most widely used, made possible by the ability to align mobility data with census-based demographic benchmarks \cite{yabe_behavioral_2023, yabe_behind_2023}. Yet despite the availability of such tools, many studies, particularly those produced during the pandemic using commercial mobility datasets, have not implemented them. As a result, their findings must be interpreted with caution: uncorrected 
demographic biases are well known to distort core mobility measures, raising important questions about the robustness of the conclusions drawn.

Compared to demographic bias, temporal bias is a less understood limitation of mobility data (see Table \ref{tab:bias_grouped}). This bias arises because GPS data generation is tied to how and when people use their phones \cite{grantz_use_2020, yoo_quality_2020, keusch_you_2022, bahr_missing_2022, wu_location-based_2024, mccool_maximum_2024}. First, location data are often collected through particular apps, meaning that mobility traces can be linked to specific behaviors (e.g., navigation, shopping, or fitness), while other everyday activities may go largely unrecorded \cite{buckee_aggregated_2020}. Second, GPS activity is uneven across the day: people are more likely to generate data at certain times when apps request location access, leading to over-representation of those periods and under-representation of others. Third, technical restrictions imposed by mobile operating systems also shape the data. For instance, location services may be limited when the battery is low, when the phone is idle, or when signals are blocked indoors, all of which can reduce the number of observations late in the day or for specific activities and places \cite{yoo_quality_2020}. Together, these mechanisms can distort measurements of mobility outcomes such as the frequency of visits to points of interest. Moreover, because patterns of app use and device behavior vary across socioeconomic and cultural groups, temporal and behavioral biases can interact with demographic bias, further skewing aggregated metrics. While a few studies have discussed aspects of temporal or activity-related sampling biases \cite{keusch_you_2022, bahr_missing_2022, yoo_quality_2020}, there is still no comprehensive effort to measure these biases systematically. Additionally, efforts to address temporal bias have primarily focused on imputation of individual-level location observations, falling short of the external validation that population-representative survey data has provided for addressing demographic bias \cite{yoo_quality_2020, hwang_comparison_2022, mccool_maximum_2024}.

\begin{table*}[t]
\centering
\small
\setlength{\tabcolsep}{6pt}
\renewcommand{\arraystretch}{1.25}
\begin{tabular}{p{2.2cm} p{3cm} p{2.5cm} p{3.3cm} p{3.2cm}}
\toprule
\textbf{Bias Type} &
\textbf{Mechanism} &
\makecell{\textbf{Key}\\\textbf{References}} &
\textbf{How Bias is Measured} &
\textbf{Correction or Validation} \\
\midrule

\multicolumn{5}{l}{\textit{I. Data \& Measurement Biases (device, sensing, coverage, panel mechanics)}} \\
\midrule

\textbf{Technical} &
Sensor errors, OS restrictions, and missing GPS fixes that distort or omit location traces. &
\cite{yoo_quality_2020, bahr_missing_2022, mccool_maximum_2024, hwang_comparison_2022, barnett_inferring_2020, beukenhorst_understanding_2021} &
Missingness and accuracy diagnostics; user-surveys. &
GPS filtering; gap imputation \cite{yoo_quality_2020, barnett_inferring_2020}; device-level calibration. \\[0.6em]

\textbf{Spatial coverage} &
Uneven data density due to signal loss, indoor / urban interference, or commercial collection focus. &
\cite{li_understanding_2024, yabe_mobile_2022, saxena_poiformer_2025}. &
Sampling density vs.\ population/land-use; under-coverage maps. &
Spatial re-weighting; integrate GPS / CDR / surveys; map- and temporal-matching \cite{saxena_poiformer_2025}. \\[0.6em]

\textbf{Selection / panel} &
Shifts in the pool of apps or users sharing data over time alter panel composition. &
\cite{safegraph_what_2019, coston_leveraging_2021, levin_insights_2021} &
Track changes in devices at home; compare active vs.\ inactive device traits. &
Longitudinal weighting; propensity models for participation. \\
\midrule

\multicolumn{5}{l}{\textit{II. Representation \& Behavior Biases (who is captured and how they use devices/places)}} \\
\midrule

\textbf{Demographic} &
Unequal phone or app use by income, age, race, or education levels. &
\cite{li_understanding_2024, Jardel2024Uncovering, wesolowski_impact_2013, coston_leveraging_2021} &
Home-CBG vs.\ census benchmarks; over- / under-representation by group. &
Pre-stratification \cite{klein_characterizing_2024}, Post-stratification \cite{moro_mobility_2021, athey_estimating_2021}; synthetic data \cite{berke_generating_2022}; validation with surveys, other datasets \cite{eagleston_features_2024,moro_mobility_2021}. \\[0.6em]

\textbf{Temporal / behavioral} &
Sampling tied to app activity; uneven pings by hour / day. Phone use differs by activity (commute vs.\ leisure) &
\cite{grantz_use_2020, yoo_quality_2020, keusch_you_2022, bahr_missing_2022, wu_location-based_2024, mccool_maximum_2024} &
User-survey data; Hourly visitation vs.\ time-use data (e.g., ATUS \cite{bls_2017}). &
GPS imputation \cite{hwang_comparison_2022, mccool_maximum_2024}; synthetic data \cite{ma_beyond_2025}; temporal re-weighting (this study). \\
\bottomrule
\end{tabular}
\caption{\textbf{Sources of bias in mobility data grouped by origin (data / measurement vs. representation / behavior) and common correction and validation approaches.}}
\label{tab:bias_grouped}
\end{table*}

In this paper, we introduce a methodology that jointly addresses temporal bias, demographic bias, and their interaction in mobility data by re-weighting behavioral measurements at population scale.  Ideally, we would do so using a reference data set analogous to the census that comprehensively captures societal patterns of activities and locations. Because such a data set does not exist, we instead leverage the American Time Use Survey (ATUS) \cite{bls_atus_home}; a high-quality, demographically-representative survey of activities in the U.S. Although not comprehensive, the ATUS provides an opportunity to benchmark behavioral patterns from mobility data against a trusted and openly-available data source. Our research has two key aims: (1) comparing the overall frequency of activities reported in ATUS to activities recorded in mobility data; and (2) re-weighting mobility insights to improve the accuracy of behavioral measurement. To demonstrate the utility of our approach, we apply our temporal re-weighting methodology to the measurement of income segregation, a major focus of recent research using mobility data in cities \cite{xu_using_2025}. Specifically, we examine the extent to which reweighting mobility data according to the ATUS impacts estimates of income segregation at specific points of interest (POIs), estimated via the method proposed by Moro et al. (2021) \cite{moro_mobility_2021}. 

We observe systematic temporal and behavioral differences—both overall and across socioeconomic groups—between activities reported in ATUS and those captured in Mobility Data (MD). Importantly, when temporal re-weighting is implemented, these discrepancies do not substantially affect static measures of segregation: mobility-based estimates of income segregation align closely with an ATUS-weighted benchmark. This validation supports the use of mobility data to capture broad patterns of income integration and isolation in cities. At the same time, our analysis uncovers significant instability in certain estimates of place-based segregation, particularly for less common POI categories and during periods of low activity in the mobility data. By leveraging ATUS to re-weight observed visits, we are able to quantify the degree of uncertainty in mobility-based segregation estimates. These results identify where conclusions drawn from mobility data are most reliable, while also clarifying the analytical limitations imposed by biases in the data-generating process.

\section*{Results}

\subsubsection*{}
\textbf{Comparing mobility visitation with official time-use estimates.} To ground our analysis, we began by examining how mobility traces reflect the everyday activities people report in population-representative surveys. This required comparing visit patterns in mobility data with time-use estimates from the ATUS. Mobility data is location-based while ATUS is activity-based. As such, ATUS has information on the type, location, start-, and end-time of each activity that respondents reported doing in one day. In contrast, mobility data infers from cellphone pings where a user was and for how long, but we do not know for certain what they were doing there. Due to these differences, we constructed a crosswalk between POI categories in the mobility data and the activities reported in the ATUS (see Methods, Relating ATUS and mobility data section). Our analysis focuses on three POI categories that corresponded to specific activities in the ATUS which were also easy to identify locations in the mobility data: Food and Coffee (Food or Food establishments), Grocery stores, and Gyms.

\begin{figure}[H]
    \centering
    \includegraphics[width=1\linewidth]{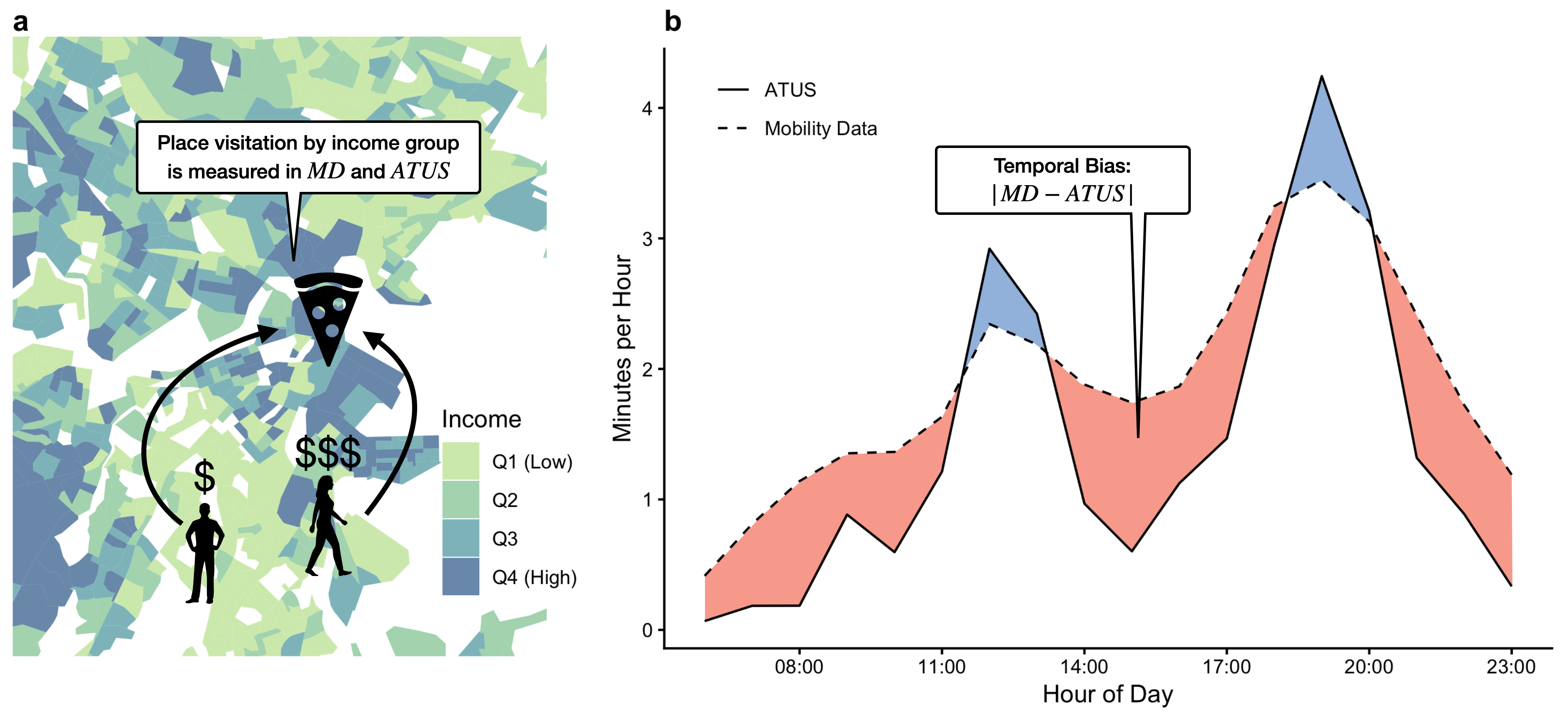}
    \captionof{figure}{\textbf{Temporal differences between ATUS and mobility data reveal temporal bias in mobility data-based time-use estimates.} \textbf{a)} Mobility data captures visitation by individuals in different income quartiles to the same POI category. \textbf{b)} Temporal bias indicates different quantities of time recorded by mobility data and ATUS for a specific POI category (shaded areas). The difference between Mobility Data and ATUS reported time indicates over- or under-representation by mobility data for individuals in a given income group. Panel b shows visitation by high income (Q4) individuals to Food \& Coffee POIs in Boston.}
    \label{fig:fig_intro}
\end{figure}

Using location data from 1.8 million devices across 11 U.S. cities between October 2016 and March 2017, we measured the amount of time individuals from different income quartiles spent at individual POIs from those categories (see Fig. \ref{fig:fig_intro}a). We then aggregated time spent within the three POI categories and compared this with time recorded in the 2017 ATUS for corresponding activities (e.g. mobility data time spent in Food and Coffee places is compared to ATUS time spent \enquote{Eating Outside the Home}). Any mismatch between time recorded by mobility data and the ATUS indicates temporal bias, where data sources offer different estimates of the timing and duration of specific activities. As we can see in Figure \ref{fig:fig_intro}b, temporal bias emerges as systematic differences between mobility- and survey-based measures across the hours of the day. 

\subsubsection*{}
\begin{figure}[H]
    \centering
    \includegraphics[width=1\linewidth]{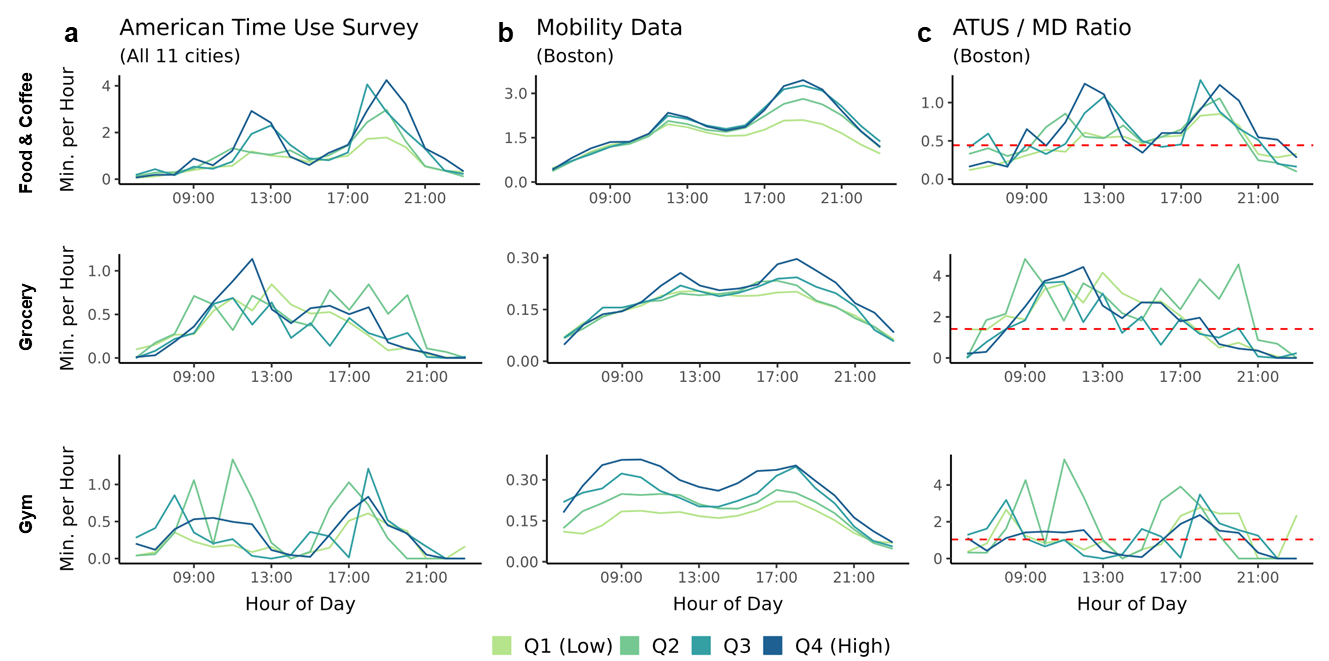}
    \captionof{figure}{\textbf{Comparison of total time spent per hour shows temporal differences between income quartiles in ATUS and mobility data between 6 a.m. and 11 p.m.} \textbf{a)} Cumulative time spent by hour for 2,502 ATUS respondents from 11 cities, by income quartile.  \textbf{b)} Cumulative time captured by hour for each income quartile from mobility data users in Boston (for results in other cities, see Supplementary Figures \ref{fig:figs1}--\ref{fig:figs3}). \textbf{c)} Ratio between ATUS and mobility data by hour for each income quartile. The red dotted line is at the mean. Hours above the mean line show when visit type is under-represented in mobility data while hours below the mean show over-representation. Hours without a ratio occur when no ATUS respondents from that quartile recorded spending time at that POI category within that hour.}
    \label{fig:fig_total_time_curves}
\end{figure}

\textbf{Temporal differences across POI category and income group.} To understand that temporal divergence better, we examined it across different POI categories over time. Figures \ref{fig:fig_total_time_curves}a and \ref{fig:fig_total_time_curves}b illustrate how visitation patterns to Food establishments, Grocery stores, and Gyms vary over the day in ATUS and mobility data. In both sources, higher-income groups (Q3–Q4) generally spend more time at these locations than lower-income groups (Q1–Q2), and all quartiles exhibit recognizable daily rhythms consistent with expected patterns of eating, shopping, and exercising. 

However, the two datasets differ markedly in smoothness. Mobility-derived curves are consistently smooth and continuous for all categories and income groups, reflecting the very large number of observations in the GPS traces. In contrast, ATUS curves exhibit substantial noise (especially for Grocery stores and Gyms) due to the much smaller number of respondents reporting these activities at specific hours. This sparsity generates visible fluctuations even when the overall pattern is similar to mobility data (see Supplementary Table \ref{tab:atusvisitationstats}).

To quantify the difference between ATUS and GPS traces, we compute a ratio $R_{h,q,cat}$ between ATUS and mobility data, which measures the extent to which mobility-based time use is over- or under-represented relative to the ATUS benchmark (see Methods, Eq. \ref{eq:atus/mdratio}). This ratio is calculated separately for each hour of the day (\textit{h}), income quartile (\textit{q}), and POI category (\textit{cat}), with results shown in Supplementary Figures \ref{fig:figs1}–\ref{fig:figs3}.

The daily visitation patterns for all three POI categories aligned with a priori expectations in both data sets. For Food establishments, ATUS and mobility data observed clear peaks in total time spent around lunch (11 a.m.--2 p.m.) and dinnertime (5--9 p.m.). Meanwhile, Grocery store visits peaked around lunch (12 p.m.) and, in the mobility data, after work (6 p.m.). And Gym visits were concentrated earlier in the morning (8--11 a.m.) and afternoon (4--6 p.m.) in both data sets. These results capture consistency between mobility data and ATUS visitation for POIs.

There was considerable variation in the ratio of minutes reported in mobility data compared to ATUS across time of day and income quartile (i.e., $R_{\:h,\:q,\:cat}$; see Fig. \ref{fig:fig_total_time_curves}c). Multiple income groups show both over- and under-representation in mobility data at a given POI category during the day (e.g., Q4 at Food establishments, Q3 at Grocery stores, and Q1 at Gyms). This indicates that, not only do individuals with different demographic characteristics visit POI categories at different times of day, but the accuracy with which their visitation is represented in mobility data also varies. This highlights the need for a technique for temporal re-weighting of mobility data that addresses the variability in how demographic groups' visitation is represented throughout the day.

\subsubsection*{}
\textbf{Re-estimating place-based income segregation weighting using official time-use estimates.} Similar to demographic post-stratification techniques, we use the values of $R_{\:h,\:q,\:cat}$ to re-weight estimates of how much time individuals from each income quartile spend during every hour of the day at each individual POI in our three POI categories of interest in the 11 cities in our sample (see Eq. \ref{eq:atus/mdratio} and the Methods section). To evaluate how much these temporal and demographic adjustments matter in practice, we apply the re-weighting procedure to a key behavioral indicator derived from mobility data: place-based income segregation.  

First, we calculate segregation with the raw number of mobility data-minutes spent by each income quartile at each POI (see Methods, Eqs. \ref{eq:tau} and \ref{eq:seg}). Then, we multiply the number of raw minutes by the ratio to obtain the re-weighted number of minutes (see Eq. \ref{eq:weighted_N}). Lastly, we repeat the same segregation calculations except with the re-weighted number of minutes to get the re-weighted segregation (see Eqs. \ref{eq:weighted_tau} and \ref{eq:weighted_seg}). The resultant segregation measures range from 0 to 1, with 0 signifying that the total time spent at a POI is evenly split across representatives from each of the four income quartiles, and 1 signifying that it is only visited by users from one income quartile \cite{moro_mobility_2021}. 

\begin{figure}[H]
    \centering
    \includegraphics[width=1\linewidth]{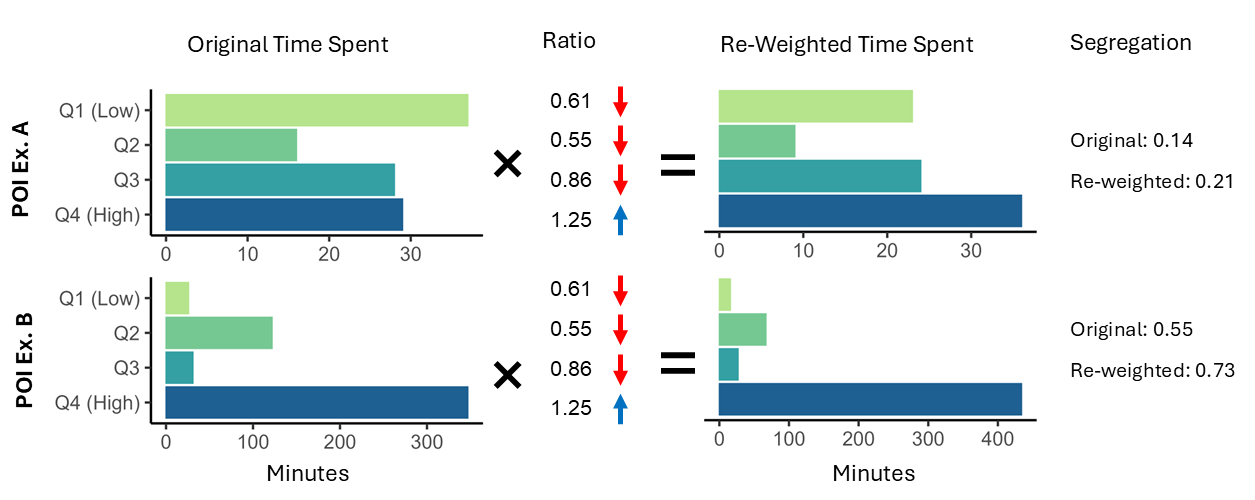}
    \caption{\textbf{ATUS / MD ratio and minutes per quartile result in smaller or larger shifts in segregation after re-weighting.} Two Food \& Coffee POIs in Boston at 12 p.m. which exhibit different changes in segregation after re-weighting.}
    \label{fig:fig_ratio_exs}
\end{figure}

Re-weighting mobility data by time-use from the ATUS has the potential to alter the composition of the income groups visiting a location, resulting in a different level of income segregation. This is demonstrated in  Figure \ref{fig:fig_ratio_exs}, in which one POI experienced a modest increase in estimated segregation after re-weighting (0.14 rising to 0.21) and another sees a more substantial shift (0.55 to 0.73). The latter example is notable because it captures a prominent consequence of the arithmetic underlying the re-weighting process. In each time period, one or more quartiles will be underrepresented in mobility data, resulting in a value of $R_{\:h,\:q,\:cat}$ that is greater than 1 and serves to substantially up-weight that category. Some POIs, however, will have robust visitation from that income quartile during that time period, meaning that an already well-represented group will be up-weighted, making it predominant after re-weighting. This translates into a greater increase in estimated segregation. In the second example in Figure \ref{fig:fig_ratio_exs}, this dynamic is visible for the fourth quartile.

\begin{figure}[H]
    \centering
    \includegraphics[width=1 \linewidth]{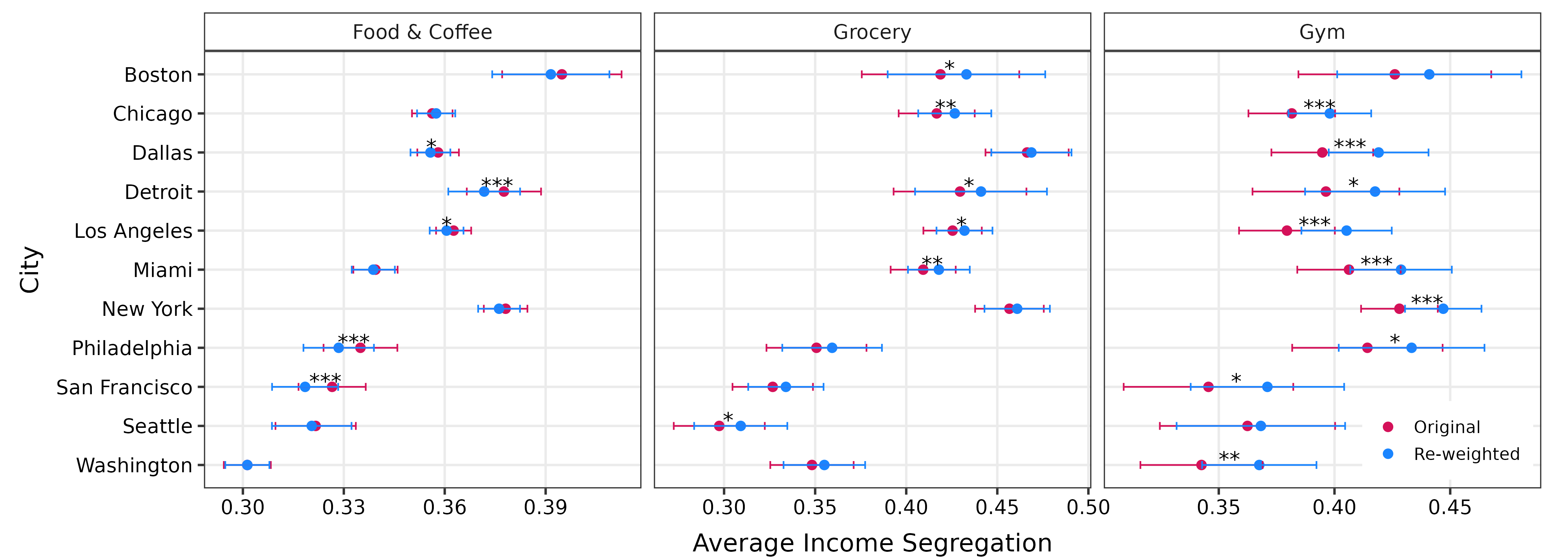}
    \captionof{figure}{\textbf{Average income segregation decreases for Food establishments but increases for Grocery stores and Gyms.} Comparing average income segregation by metro-area before and after re-weighting with ATUS / MD ratios for each POI category. Significance codes: 0 '***' 0.001 '**' 0.01 '*' 0.05}
    \label{fig:fig_ave_inc_seg}
\end{figure}

The re-estimation of segregation at individual POIs could consequently alter the estimated degree of income segregation throughout a given city. When averaged across POI categories and time, however, we find relative agreement between citywide averages in segregation derived from the original and re-weighted data. The lowest average absolute effect of re-weighting on segregation was in Food POIs $\abs{\Delta S_{\alpha}}=0.003$ (95\% CI [0.001, 0.005]), followed closely by Grocery stores $\abs{\Delta S_{\alpha}}=0.008$ (95\% CI [0.006, 0.011]). The highest was in Gyms $\abs{\Delta S_{\alpha}}=0.02$ (95\% CI [0.016, 0.024]). Further, re-weighting maintained both city- and POI category-specific differences in segregation (see Fig. \ref{fig:fig_ave_inc_seg}). This was also confirmed with two-way ANOVAs which resulted in \textit{p}-values less than 0.001 for city and POI category for original and re-weighted segregation. 

Across cities, the direction of change in income segregation varied by POI category. Average income segregation after re-weighting was lower for Food establishments in 9 of 11 cities, but at Grocery stores and Gyms it was higher after re-weighting in all cities.
These differences were significant in the majority of cases: 5/11 cities (45\%) for Food establishments, 6/11 (55\%) for Grocery stores, and 9/11 (82\%) for Gyms, though given the modest shifts this is primarily a reflection of large sample sizes. 

These results are promising for the use of mobility data to measure income segregation, as adjustments for potential temporal bias based on the ATUS produced marginal differences in the overall estimate of segregation across POIs. This adds nuance to the role played by ATUS re-weighting, showing that the impacts of temporal bias are not the same across activity categories.

\subsubsection*{}
\begin{figure}[H]
    \centering
    \includegraphics[width=1 \linewidth]{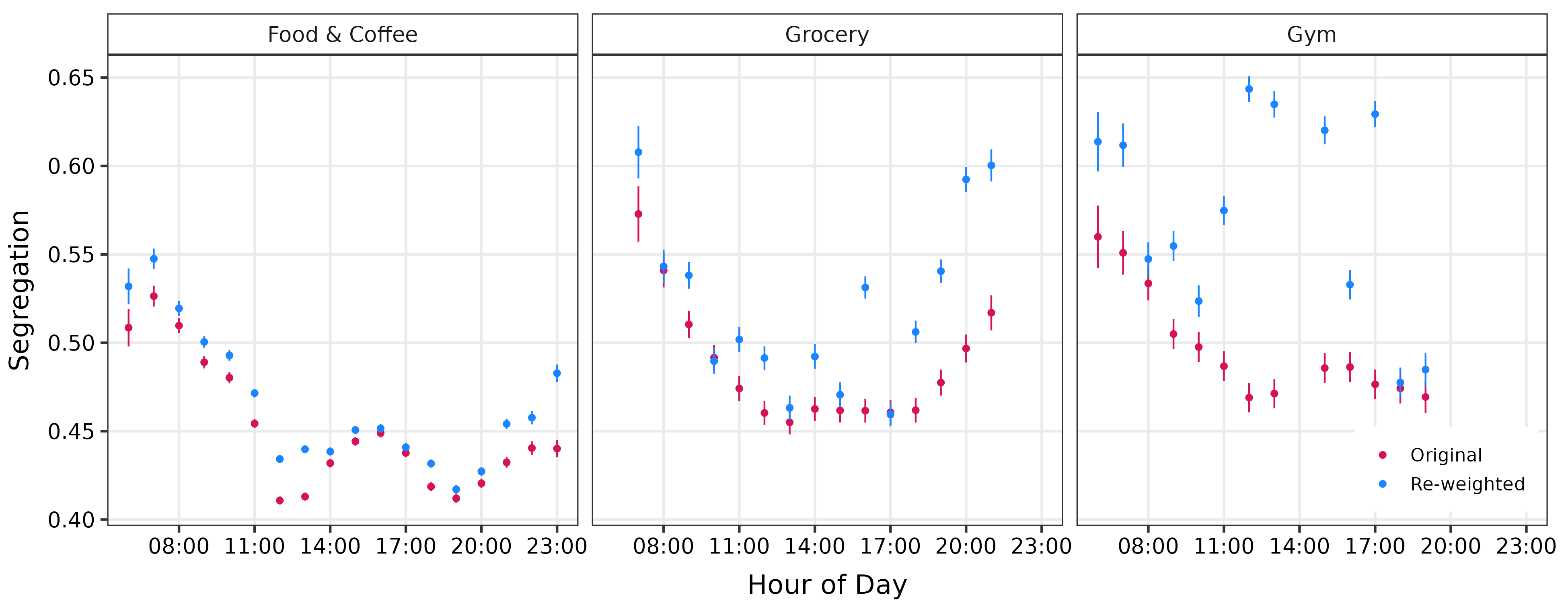}
    \captionof{figure}{\textbf{Magnitude of difference between income segregation pre- and post- ATUS-weighting throughout the day.} Hourly comparison of original and re-weighted income segregation, averaged across POIs from all 11 study cities.}
    \label{fig:fig_hourly_ave_inc_seg}
\end{figure}

\textbf{Differences in re-weighted segregation vary across time.} A major advantage of a technique accounting for temporal bias is the ability to apply it to patterns in time use across the hours of the day (see Fig. \ref{fig:fig_hourly_ave_inc_seg}). We do this here, though we first had to apply new inclusion criteria for POIs for each hour of the day, including only those that had at least 5 unique mobility data visitors (from any income quartile) at that hour. Additionally, we set a minimum of 5 ATUS respondents at that location category at that hour (see Methods section). We compared re-weighted and original income segregation by hour using paired t-tests, finding significant differences (p-value < 0.05) for all hours (6 a.m.--11 p.m.) at Food establishments, all hours (6 a.m.--1 p.m. and 3 p.m.--7 p.m.) at Gyms, and for most hours at Grocery POIs (7 a.m.--9 p.m., except for the hours of 8 a.m., 10 a.m., and 5 p.m.). 

Food establishments exhibited a clear diurnal pattern that is consistent in both original and re-weighted data. They have lower segregation at high visitation times, around lunch (12--2 p.m.) and dinner (5--8 p.m.) and the highest segregation early in the morning and late at night. Differences between original and re-weighted segregation were consistently modest except late at night, though they were often statistically significant across the day. Grocery POIs had similar diurnal patterns, with the highest segregation early in the morning and late at night, though with greater discrepancies between original and re-weighted estimates, especially in the afternoon (12--6 p.m.). Meanwhile, Gyms exhibited the largest discrepancies between original and re-weighted segregation, with re-weighted data describing much higher levels of segregation from 12 p.m. to 5 p.m.  (see Fig. \ref{fig:fig_hourly_ave_inc_seg}).

As noted, all three POI categories had certain  hours with a more pronounced difference between original and re-weighted segregation. This may be an aggregated version of the same dynamic captured in Figure \ref{fig:fig_ratio_exs} at the POI level. In each time period, one or more quartiles may be substantially up-weighted by our technique. If certain POIs are heavily visited by individuals from that income quartile during that time, then its estimated segregation will increase with re-weighting. If this is true for numerous POIs, then estimated average segregation will increase. This could explain, for example, increases in segregation in Gyms during the day, as representatives of quartile four are under-represented in mobility data relative to ATUS, but many establishments have a majority of visitors from that quartile.

Another potential driver of these differences is the volume of visitors. Because the beginning and end of the day have the fewest unique visitors, re-weighting is liable to have greater impacts on estimates, whereas re-weighting based on larger numbers will be more robust. We see this for late mornings, afternoons, and earlier part of the evenings for Food establishments and Grocery stores (see Supplemental Information, Fig. \ref{fig:addendumfig4}). In sum, periods with greater overall representation in mobility data and ATUS see less discrepancy between income segregation across categories.  

\subsubsection*{}
\textbf{Modeling differences in segregation.} 
Accounting for temporal and demographic bias in mobility data produced very similar average levels of segregation across the POIs of a given city. Nonetheless, there was substantial variation in the extent to which estimates of segregation at individual POIs shifted. For instance, while 43\% of POIs saw either an increase or decrease in segregation of .05 or greater, only 11\% saw an increase or decrease of .1 or greater, with a mean deviation of .05 across all POIs. This is especially relevant for studies that represent and analyze flow dynamics at individual POIs \cite{moro_mobility_2021, yabe_behavioral_2023, athey_estimating_2021, xu_using_2025}. To better understand why re-weighting adjusts segregation estimates more at some POIs than others, we ran multiple linear regressions predicting shifts in the estimates using: (1) number of POIs per census block group (CBG), a proxy for POI density which may cause misattribution of visits to POIs; (2) number of unique visitors to a POI, a proxy for POI popularity; and (3) the fraction of users by income quartile, proxies for whether POIs predominantly serve specific income segments, potentially indicating that POIs targeting particular ends of the income spectrum are especially sensitive to re-weighting (see Table \ref{tab:mult_lin_reg}). All models account for city-level variation with fixed effects. That said, the conditional and marginal $R^2$ values were equal for all three models, indicating that cities explained no additional variance. 

There was a significant and negative association between original segregation and the change in segregation for all three POI categories (Food: \textit{p} < .001, $\beta$ = -0.17; Grocery: \textit{p} < .001, $\beta$ = -0.24; Gym: \textit{p} < .001, $\beta$ = -0.27), likely due to the fact that when original segregation is higher, re-weighting is more likely to result in a lower estimate of segregation. This is in part an artifact of ceiling effects; when an estimate of segregation is very high, it is arithmetically difficult for it to go higher. Meanwhile, the number of POIs per CBG was not significantly associated with change in the estimate of segregation for any category (Food: \textit{p} = .16, $\beta$ = 0.01; Grocery: \textit{p} = .72, $\beta$ = -0.01; Gym: \textit{p} < .88, $\beta$ = 0), which suggests that the algorithm used to assign users to nearby POIs for these three categories is robust (see Methods section).

The number of unique visitors (Food: \textit{p} < .001, $\beta$ = -0.03; Grocery: \textit{p} < .001, $\beta$ = -0.06; Gym: \textit{p} = .05, $\beta$ = -0.04) and the fraction of Q1 visitors (Food: \textit{p} < .001, $\beta$ = -0.42; Grocery: \textit{p} < .001, $\beta$ = -0.12; Gym: \textit{p} < .001, $\beta$ = -0.3) both predicted lower estimates of segregation after re-weighting for all three POI categories. In contrast, the fraction of Q4 visitors to a POI predicted an upward shift in estimated segregation for Food and Grocery POIs but a downward shift in estimated segregation for Gyms (Food: \textit{p} < .001, $\beta$ = 0.17; Grocery: \textit{p} < .001, $\beta$ = 0.1; Gym: \textit{p} < .001, $\beta$ = -0.44). These seemingly contradictory results likely reflect distinctions in the ratios that underlie re-weighting. Food and Grocery POIs tend to see a modest under-representation of people from higher income communities, meaning that visits by Q4 individuals are up-weighted. Thus, places that already have a high proportion of Q4 visitors will be re-weighted to be more segregated (i.e., even more Q4 visitors relative to other groups). Gyms, however, have more visitors from Q4 and consequently are down-weighted, meaning segregation at places dominated by Q4 visitors is tempered in the process. The models for Food and Gyms explained 35\% and 23\% of the variance, respectively, whereas the model for Grocery stores was the least explanatory at 12\%.

\begin{table}[ht]
\centering
\begin{tabular}{|lllllll|} \hline
\multirow{2}{*}{\textbf{Explanatory Variable}} & \multicolumn{2}{c}{\textbf{\emph{Food $\&$ Coffee}}} & \multicolumn{2}{c}{\textbf{\emph{Grocery}}} & \multicolumn{2}{c|}{\textbf{\emph{Gym}}} \\ \cline{2-7} &
 \textbf{Std. $\beta$} & \textbf{Std. Error} & \textbf{Std. $\beta$} & \textbf{Std. Error} & \textbf{Std. $\beta$} & \textbf{Std. Error} \\[1pt] 
\noalign{\hrule height 1pt} 
\rule{0pt}{2.5ex}Segregation             & -0.17 *** & 0.01 & -0.24 *** & 0.02 & -0.27 *** & 0.02 \\
$\#$ POIs per CBG       & $ $ 0.01     & 0    & -0.01     & 0.02 &  $ $ 0        & 0.02 \\
$\#$ Unique visitors    & -0.03 *** & 0.01 & -0.06 *** & 0.02 & -0.04 *   & 0.02 \\
Fraction of Q1 visitors & -0.42 *** & 0.01 & -0.12 *** & 0.02 & -0.3  *** & 0.02 \\
Fraction of Q4 visitors & $ $ 0.17 *** &  0.01 & $ $ 0.1 $ $*** & 0.02 & -0.44 *** & 0.02 \\
\noalign{\hrule height 1pt}
\multicolumn{1}{|l}{\rule{0pt}{2.5ex}\textbf{Model Results}} & 
\multicolumn{2}{c}{} &
\multicolumn{2}{c}{} &
\multicolumn{2}{c|}{}
 \\[1pt] 
\noalign{\hrule height 1pt}
\rule{0pt}{2.5ex}Conditional R$^2$ & \multicolumn{2}{c}{0.35} & \multicolumn{2}{c}{0.12} & \multicolumn{2}{c|}{0.23} \\
Marginal R$^2$ & 
\multicolumn{2}{c}{0.35} & \multicolumn{2}{c}{0.12} & \multicolumn{2}{c|}{0.23} \\ \hline
\end{tabular}
\caption{\textbf{Number of unique mobility data visitors along with the fraction of visitors from quartiles 1 (lower income) and 4 (higher income) have significant relationships with the change in segregation after re-weighting.} Multiple linear regression results by POI category across all 11 cities. The dependent variable for these models is the difference between re-weighted and original segregation \(S_\alpha^{\:wt.} - S_\alpha^{\:orig.}\). Significance codes: 0'***'0.001'**'0.01"*'0.05 }
\label{tab:mult_lin_reg}
\end{table}

\section*{Discussion}

This study contributes to growing literature that investigates the validity and limits of large-scale mobility data as a proxy for human behavior in cities. Mobility data can reproduce aggregated spatial patterns of visitation, interaction, and segregation \cite{athey_estimating_2021,xu_using_2025,moro_mobility_2021,tizzoni_use_2014,wesolowski_connecting_2016,buckee_aggregated_2020}, particularly when demographic biases are addressed through post-stratification or related techniques \cite{moro_mobility_2021,yabe_enhancing_2024,cabrera_systematic_2025,coston_leveraging_2021,safegraph_what_2019,aleta_modelling_2020}. However, a parallel body of work has documented that mobility data are generated through uneven, behavioral mediated sampling processes tied to phone use, app activity, and technical constrains \cite{yoo_quality_2020,keusch_you_2022,bahr_missing_2022,wu_location-based_2024,hwang_comparison_2022,saxena_poiformer_2025}. Our results help bridge these two strands of the literature by surfacing temporal bias as a distinct but unexplored dimension of representativeness in mobility data. We find that this bias emerges from the interaction of technical constraints and behaviorally mediated sampling processes and demonstrate that it can be systematically measured, validated, and partially mitigated using official time-use surveys (see Table \ref{tab:bias_grouped}).

A central finding of our work is that mobility data reproduce broad, city-level patterns of experienced income segregation with a high degree of stability, even after adjusting for temporal and demographic differences using the American Time Use Survey (ATUS) \cite{bls_atus_home}. On average, re-weighting mobility-based visitation by ATUS time-use estimates produces only modest changes in aggregate segregation levels across cities and POI categories. This consistency aligns with prior mobility-based studies of segregation \cite{moro_mobility_2021,xu_using_2025,athey_estimating_2021}, indicating that their conclusions are not driven by systematic distortions in the timing of observed activity. At the same time, the limited magnitude of average change should not be interpreted as evidence that temporal bias is inconsequential. Rather, it reflects the fact that segregation—being an aggregated, compositional measure—is relatively robust when averaged across time and space, even as substantial variation emerges at finer temporal and spatial resolutions. Differences between original and re-weighted segregation are most pronounced at specific times of day, particularly during early morning and late evening hours when device coverage is lower. It is also much more noteworthy at individual POIs, especially those with small numbers of visitors or whose visitors are predominantly from a single income bracket. As shown in Table \ref{tab:mult_lin_reg}, shifts in segregation after re-weighting depend systematically on POI characteristics, including the number of unique visitors and the income distribution of those visitors, as well as on activity type. These patterns reflect the mechanics of temporal bias and the re-weighting, whereby under-represented income groups at particular hours are scaled upward, amplifying or attenuating segregation depending on existing visitation structure. Differences across POI categories, including the greater variability observed for Gyms relative to Food establishments, are consistent with differences in activity frequency and temporal concentration.

Although income segregation serves as the empirical focus of this analysis, the implications extend to other mobility-based applications. Segregation metrics depend on relative composition and are sensitive to sparse observations, making them a demanding test case for assessing temporal bias. Applications that rely more directly on time-of-day patterns may be even more affected. For example, studies of commuting behavior or shift work often focus on early morning or late-night hours, when mobility data coverage is lowest and participation varies systematically by occupation and income, potentially biasing estimates of exposure, accessibility, or travel disruption \cite{wang_mitigating_2025,eagleston_features_2024}. In disaster contexts, analyses of evacuation, sheltering, or recovery frequently rely on short temporal windows immediately before or after an event, when changes in phone use, battery depletion, or network outages can distort observed mobility patterns \cite{kissler_reductions_2020,yabe_mobile_2022}. Similarly, analyses of leisure activity, nightlife, or policy interventions tied to specific hours may be influenced by uneven behavioral sampling across the day \cite{aleta_quantifying_2022,buckee_aggregated_2020}.  In these settings, unexamined temporal bias can alter comparisons across groups, places, or time periods by conflating behavioral change with shifts in data coverage.

A limitation of the proposed temporal re-weighting procedure is its reliance on official time-use estimates, which are not widely available and are typically based on relatively small samples. In the case of the American Time Use Survey (ATUS), approximately 10,000 respondents are surveyed annually in the United States, with responses re-weighted to represent national rather than city-specific time-use patterns, constraining reliability for less frequent activities and fine temporal disaggregation. These constraints are particularly apparent for activities such as exercising in gyms, which occur infrequently in both ATUS and mobility data and exhibit greater instability when disaggregated by hour and income group. Additional limitations arise in aligning activity-based survey data with location-based mobility data, as time-use surveys provide limited spatial detail beyond broad location categories, restricting the range of POI types that can be meaningfully analyzed. As a result, temporal re-weighting is best viewed as a complementary diagnostic and mitigation approach, rather than a standalone solution, alongside demographic post-stratification and validation using independent data sources.

Despite the widespread use of mobility data in urban research, relatively few studies systematically compare mobility-derived measures with official statistical sources or implement explicit post-stratification and related bias-mitigation procedures. Without such comparisons, it is difficult to distinguish urban patterns from artifacts of observation. Incorporating validation against independent benchmarks, alongside demographic and temporal adjustment strategies, provides a clearer basis for inference by ensuring that conclusions are anchored in behavior rather than in the mechanics of data collection.

\section*{Methods}
\subsubsection*{}
\textbf{Mobility data.} We analyzed anonymized, high-resolution mobile location data from Cuebiq collected between October 2016 and March 2017 across 11 cities in the U.S.: Boston, Chicago, Dallas, Detroit, Los Angeles, Miami, New York, Philadelphia, San Francisco, Seattle, and Washington (see Table \ref{tab:cbsastats}). Mobility data begins as GPS points which are clustered using the Hariharan \& Toyama algorithm \cite{hariharan_toyama_algorithm_2004}, creating a set of 'stays' with a beginning and end time. These stays are then assigned to the closest POI within a radius of 100 meters. Time spent at specific POI categories is then aggregated to describe group-level visit durations. For more information about mobility data pre-processing and details on the representativeness of the data and robustness of the pre-processing pipeline, see the Supplementary Information of Moro et al. (2021) \cite{moro_mobility_2021} and Garcia Bulle Bueno, et al. (2024) \cite{ff_visits_2024}.

To reduce mobility data stays that could have occurred at the user's estimated home location or stays where we have less certainty of the assigned POI (e.g. when POIs are close together or in the same building like shopping centers, a common occurrence in cities), we applied two distance filters. Each stay had to occur more than 50 meters from the user's estimated home location and less than 10 meters from the assigned POI location. To address and manage mobility data visits that are more likely to be work-related, we excluded visits that were longer than 4 hours. To prevent attribution of users who passed by, but did not visit a POI, visits also had to be a minimum of 5 minutes. For the analyses that we performed at the POI-level, POIs had to be visited more than 100 times and by more than 20 unique users throughout the 6-month data collection period \cite{moro_mobility_2021}.

\subsubsection*{}
\textbf{American Time Use Survey data.} ATUS data is collected, processed, and published annually by the Bureau of Labor and Statistics \cite{bls_atus_home}. Using computer-assisted telephone interviews, ATUS records 24-hours of activities for a representative sample of U.S. residents. These activities have categorical descriptions such as, "work, main job" and "eating and drinking." When provided by the respondent, there is also a general location attached to each activity such as "respondent's home or yard," "grocery store," or "car, truck, or motorcycle (driver)." ATUS intentionally does not collect locations for certain activities including: sleeping, personal / private activities, or those with insufficient or missing detail \cite{bls_atus_dd_int_2017}. To facilitate recall, respondents report their activities the day after their diary day \cite{alwin_twenty-four_2009}. 

We analyzed the 2017 ATUS data to ensure it was aligned with the collection period of our mobility data \cite{bls_2017}. We filtered respondents to those from the same 11 cities so that we compared mobility data stays to activities from ATUS respondents in major cities and excluded those in rural regions. Since the sample size for ATUS respondents in the 11 cities was relatively small (n = 2,502), the same set of ATUS respondents were compared against the mobility data from each city and were not subset further. See the Supplementary Information for more detail on the mobility data stays before and after filtering and basic statistics on mobility data visitation by POI category (Tables \ref{tab:cbsastats}, \ref{tab:bostonMDvisitationstats}).

As the ATUS aims to reflect the U.S. population at the time of each survey, a weight is calculated for each respondent to account and adjust for three major aspects of ATUS sampling: (1) as ATUS is based on a stratified sample, the weights ensure each group is correctly represented in the population, (2) as ATUS is not distributed evenly by diary day, the weights were calculated so that each day of the week is correctly represented for the sample quarter, and (3) the weights account for the fact that response rates differ across demographic groups and weekdays as compared to weekends \cite{bls_atus_users_guide_2017}. These weights were applied in our estimation of the time spent by ATUS respondents in different activities (Eq. \ref{eq:weights}),

\begin{equation} \label{eq:weights}
    \frac{\sum\limits_{i\epsilon q} W_i\tau_{i\alpha} (h)}{\sum\limits_{i\epsilon q} W_i\times 60}
\end{equation}
where \textit{i} is the ATUS individual respondent, \textit{q} is income quartile, \textit{W} is the assigned ATUS weight \cite{bls_atus_users_guide_2017}, \textit{$\alpha$} is the POI category, and \textit{h} is the hour.  

\subsubsection*{}
\textbf{Relating ATUS and mobility data.} To compare visitation and time spent between mobility data and ATUS we chose three location types that could be matched between the two data sources (see Tables \ref{tab:atusvisitationstats} - \ref{tab:bostonMDvisitationstats}). For example, one of the most common mobility POI categories is "Food \& Coffee," however, there is not a direct match in ATUS for visiting a Restaurant or Café. Instead, we calculated a comparable measure of time spent at these locations using a combination of activity and location filters in ATUS data: "eating and drinking" or "purchasing food (not groceries)" that took place at a "restaurant/bar", "other store / mall" or "other place." For Grocery and Gym visitation we were able to connect time spent more easily across datasets because ATUS records both "grocery store" and "gym / health club" locations. We used all the  ATUS activities that were recorded at those locations except for work-related activity including: working main or other jobs, and waiting or traveling related to working. To alight ATUS estimates with mobility data, we also applied the same time limit and filtered out activities lasting longer than 4 hours. Our mobility data minimum length of time is 5 minutes, which aligns across the data sets, as the shortest length of a single activity that ATUS publishes is also 5 minutes. A summary of 2017 ATUS time spent after the activity length filtering is available in the Supplementary Information (see Table \ref{tab:atusvisitationstats}).     

\subsubsection*{}
\textbf{Income quartiles.} For devices in mobility data, income quartiles were defined by the distribution of median incomes within cities, and individuals were assigned the income quartile of their residence census block group according to the procedure presented in Moro et al. (2021) \cite{moro_mobility_2021} (See Moro et al., 2021, Supplementary Figure 2). For ATUS respondents, income was determined based on each respondent's self-reported family income, and income brackets were matched to quartiles defined for mobility data. Quartile one has 1,031 ATUS respondents (41\%), quartile two has 349 respondents (14\%), quartile three has 373 respondents (15\%), and quartile four has 749 respondents (30\%). Table \ref{tab:atusincomebreaks} includes a detailed breakdown of the number of respondents per quartile for each city, their corresponding ATUS household income range, and the mobility data median income range per quartile per city \cite{moro_mobility_2021}.

\subsubsection*{}
\textbf{Quantifying differences between MD and ATUS.} We compare the cumulative time spent by ATUS respondents of each income quartile across our three POI categories (see Fig. \ref{fig:fig_total_time_curves}a) to the cumulative time spent by mobility data users of each income quartile, which we illustrate using the Boston metro-area (see Fig. \ref{fig:fig_total_time_curves}b). We then we calculate the MD / ATUS ratio for each of the 11 cities per hour, per quartile for each studied POI category (see Figs. \ref{fig:figs1}--\ref{fig:figs3}) (Eq. \ref{eq:atus/mdratio}):

\begin{equation} \label{eq:atus/mdratio}
    R_{\:h,\:q,\:cat} = 
    \cfrac{\;N_{cat}^{\:ATUS\bigstrut[b]} (h;q)\; / N_{total\bigstrut[b]}^{\:ATUS\bigstrut[b]} (h;q)}{N_{cat}^{\:MD\bigstrut[b]} (h;q)\; / N_{total}^{\:MD\bigstrut[b]} (h;q)}
\end{equation}

where \textit{N} is the number of observed minutes (normalized), for each of the three POI categories (\textit{cat}): (1) Food and Coffee establishments, (2) Grocery stores, and (3) Gyms, for each hour (\textit{h}) and income quartile (\textit{q}). The numerator equals the proportion of reported ATUS minutes out of the total number of available ATUS minutes in that hour. While the denominator equals the proportion of mobility data minutes out of the total number of available mobility data minutes recorded in that hour.  

\subsubsection*{}
\textbf{Income segregation.}
In this work we explore one application of this ratio, income segregation \cite{moro_mobility_2021}, by comparing how it affects POI-level segregation before and after the ratio is applied to re-weight the mobility data. 

Income segregation is based on the time $\tau_{\:\alpha,\:q}$ spent by individuals of different income quartiles $q$ at a specific POI $\alpha$: 

\begin{equation} \label{eq:tau}
\tau_{\:\alpha,\:q} = \;\;\cfrac { \sum\limits_{h} {N^{MD\bigstrut[b]}_{\alpha\bigstrut[b]}(h;q)}}{ \sum\limits_{q}  \sum\limits_{h} {N^{MD\bigstrut[b]}_{\alpha\bigstrut[b]}(h;q)}}
\end{equation}

Where income segregation $S_\alpha$ measures the relative balance of time spent by individuals from different income quartiles:

\begin{equation} \label{eq:seg}
    S_\alpha = \frac{2}{3} \sum_q \abs{\tau_{\:\alpha,\:q} - \frac{1}{4}}
\end{equation}

Re-weighted POI-level income segregation was calculated based on re-weighted minutes spent by a given income group at a given POI, $\hat{N}_{\alpha} (h;q)$, based on the ratio between MD and ATUS time use, $R_{\:h,\:q,\:cat}$ for each hour, quartile, and POI type computed according to Eq. \ref{eq:atus/mdratio}:

\begin{equation} \label{eq:weighted_N}
    \hat{N}_{\alpha} (h;q) = N^{MD}_{\alpha} \times R_{cat} (h;q)
\end{equation}

The re-weighted number of mobility data minutes was then used to calculate a re-weighted measure of time spent at a given POI by each income quartile, $\hat{\tau}_{\:\alpha,\:q}$:

\begin{equation} \label{eq:weighted_tau}
\hat{\tau}_{\:\alpha,\:q} = \;\;\cfrac { \sum\limits_{h} {\hat{N}^{MD\bigstrut[b]}_{\alpha\bigstrut[b]}(h;q)}}{ \sum\limits_{q}  \sum\limits_{h} {\hat{N}^{MD\bigstrut[b]}_{\alpha\bigstrut[b]}(h;q)}}
\end{equation}

Finally, re-weighted segregation $\hat{S}_\alpha$ was computed using re-weighted time use estimates per income quartile:
\begin{equation} \label{eq:weighted_seg}
    \hat{S}_\alpha = \frac{2}{3} \sum_q \abs{\hat{\tau}_{\:\alpha,\:q} - \frac{1}{4}}
\end{equation}

where \(\tau_{\alpha,q}\)  represents the proportion of time at place $\alpha$ spent by income quartile \textit{q}. POI-level segregation is quantified as a measure between 0 and 1, where \(S_\alpha = 0\) is a fully income integrated POI signifying that the total time spent at that venue is evenly split across users from each of the four income quartiles. And where \(S_\alpha = 1\) is a fully income segregated POI, signifying that it is only visited by users from one income quartile \cite{moro_mobility_2021}. 

\subsubsection*{}
\textbf{Multiple linear regression.} To better understand the factors associated with differences between re-weighted and original segregation we used a multiple linear regression model. The independent variables for this model were: segregation, number of POIs per census block group, number of unique visitors, fraction of quartile 1 visitors, and fraction of quartile 4 visitors. While running this regression, metro-area was taken into account by setting city to a factored dependent variable. We constrained the regression to POIs with an original income segregation less than or equal to 0.95 to prevent extremely segregated POIs from skewing the results. We added additional filtering to ensure that the POIs and hours included for Figure \ref{fig:fig_hourly_ave_inc_seg} received at least 5 unique visitors during that hour over the course of the 6-month mobility data period, and that at least 5 ATUS respondents recorded doing that activity at that hour. 

\section*{Data Availability} American Time Use Survey data is publicly available from the Bureau of Labor and Statistics \cite{bls_2017}. Mobility data are available by application to Cuebiq through their Social Impact program: \href{ https://cuebiq.com/social-impact/}{https://cuebiq.com/social-impact/}. Mobility data were used under license for the current study. Data and code to reproduce the main results of this analysis are publicly available under MIT license from: \href{https://github.com/SUNLab-NetSI/mobility_data_temporal_bias}{https://github.com/SUNLab-NetSI/mobility\_data\_temporal\_bias}.

\clearpage
\bibliography{references}

\clearpage
\section*{Supplementary Information}

\renewcommand{\thetable}{S\arabic{table}}
\renewcommand{\thefigure}{S\arabic{figure}}
\setcounter{equation}{0}
\setcounter{table}{0}
\setcounter{figure}{0}

\subsection*{Mobility Data}

\begin{table}[H]
    \centering
    \begin{tabular}{|llllll|} \hline
      \multirow{2}{*}{\textbf{City (State)}} &
      \multirow{2}{*}{\textbf{Population}} &
      \multicolumn{2}{c}{\textbf{\emph{Raw Data}}} & \multicolumn{2}{c|}{\textbf{\emph{Dist. Filtered}}}\\  \cline{3-6}
    & & \textbf{\# Devices} & \textbf{\# Stays} & \textbf{\# Devices} & \textbf{\# Stays} \\ \hline
    Boston-Cambridge-Newton (MA-NH) & 4.64 M & 74 k & 12.35 M & 67 k & 1.35 M\\
    Chicago-Naperville-Elgin (IL-IN-WI) & 9.52 M & 304 k & 68.16 M & 281 k & 5.72 M\\
    Dallas-Fort Worth-Arlington (TX) & 6.70 M & 246 k & 55.02 M & 231 k & 4.73 M\\
    Detroit-Warren-Dearborn (MI) & 4.29 M & 128 k & 26.79 M & 117 k & 2.13 M\\
    Los Angeles-Long Beach-Anaheim (CA) & 13.05 M & 244 k & 63.3 M & 237 k & 6.98 M\\
    Miami-Fort Lauderdale-West Palm Beach (FL) & 5.76 M & 187 k & 46.22 M & 180 k & 4.42 M\\
    New York-Jersey City (NY-NJ-PA) & 19.83 M & 248 k & 58.42 M & 238 k & 7.5 M\\
    Philadelphia-Camden-Wilmington (PA-NJ-DE-MD) & 6.02 M & 123 k & 23.91 M & 115 k & 2.17 M\\
    San Francisco-Oakland-Hayward (CA) & 4.46 M & 86 k & 17.38 M & 83 k & 1.91 M\\
    Seattle-Tacoma-Bellevue (WA) & 3.55 M & 73 k & 14.19 M & 70 k & 1.43 M\\
    Washington-Arlington-Alexandria (DC-VA-MD-WV) & 5.86 M & 153 k & 28.83 M & 144 k & 2.84 M\\ \hline
    \textbf{Total} & 83.5 M & 1.87 M & 415 M & 1.76 M & 41.2 M\\ \hline
        \end{tabular}
    \caption{Mobility data set information for the 11 cities in this study, including the raw number of devices and stays and the number of devices and stays after applying distance filters: more than 50 meters from the user's estimated home location and less than 10 meters from the assigned Foursquare POI location \cite{moro_mobility_2021}.}
    \label{tab:cbsastats}
\end{table}

\subsection*{ATUS}
Table \ref{tab:atusincomebreaks} shows how we assigned income quartiles (1-4) to ATUS respondents. Respondents from each city were split to best match the income quartile splits assigned to each unique mobility data user \cite{moro_mobility_2021}, explaining why the mobility data assignments are much closer to true quartiles and why ATUS assignments are significantly further away (Tables \ref{tab:atusvisitationstats}--\ref{tab:bostonMDvisitationstats}).

\begin{table}[H]
    \centering
    \begin{tabular}{|lcccc|} \hline
    \multirow{2}{*}{\textbf{City}} &
    \multirow{2}{*}{\textbf{Income Quartile}} &
    \multicolumn{2}{c}{\textbf{\emph{ATUS}}} & \multicolumn{1}{c|}{\textbf{\emph{Mobility Data}}}\\ \cline{3-5}
    & & \textbf{Respondents (\#)} & \textbf{Family Income (\$)} & \textbf{Median Income (\$)}\\ \hline
    \multirow{4}{*}{Boston} & 1& 40& $\leq$ 59,999& $\leq$ 66,776\\
    & 2& 40& 60,000 - 99,000 & 66,777 - 90,368\\ 
    & 3& 45& 100,000 - 149,000 & 90,369 - 114,167\\ 
    & 4& 41& 150,000 + & 114,168 +\\ \hline
    \multirow{4}{*}{Chicago} & 1& 65& $\leq$ 49,000 & $\leq$53,631\\
    & 2& 94& 50,000 - 74,999 & 73,583\\
    & 3& 91& 75,000 - 99,999 & 97,371\\
    & 4& 46& 100,000 + & 97,372 +\\ \hline
    \multirow{4}{*}{Dallas} & 1& 65& $\leq$ 49,999 & $\leq$ 48,594\\
    & 2& 68& 50,000 - 74,999 & 48,595 - 71,081\\
    & 3& 56& 75,000 - 99,999 & 71,082 - 97,292\\
    & 4& 26& 100,000 + & 97,293 +\\ \hline
    \multirow{4}{*}{Detroit} & 1& 39& $\leq$ 39,999 & $\leq$ 46,471\\ 
    & 2& 42& 40,000 - 59,999& 64,514\\
    & 3& 28& 60,000 - 74,999& 88,750\\
    & 4& 16& 100,000 + & 88,751+\\ \hline
    \multirow{4}{*}{Los Angeles} & 1& 94& $\leq$ 49,999 & $\leq$ 47,500\\
    & 2& 98& 50,000 - 74,999 & 47,501 - 68,500\\
    & 3& 83& 75,000 - 99,999 & 68,501 - 94,167\\
    & 4& 47& 100,000 + & 94,168 +\\ \hline
    \multirow{4}{*}{Miami} & 1& 72& $\leq$ 39,999 & $\leq$ 41,563 \\
    & 2& 68& 40,000 - 59,999 & 41,564 - 59,090\\
    & 3& 28& 60,000 - 99,999 & 59,091 - 82,405\\
    & 4& 16& 100,000 + & 82,406 +\\ \hline
    \multirow{4}{*}{New York} & 1& 146& $\leq$ 59,999 & 61,115\\
    & 2& 131& 60,000 - 74,999 & 61,116 - 86,591\\
    & 3&152& 75,000 - 99,999 & 86,592 - 116,369\\
    & 4& 111& 100,000 + & 116,370 +\\ \hline
    \multirow{4}{*}{Philadelphia} & 1& 43& $\leq$ 49,999 &  $\leq$ 55,313\\
    & 2& 52& 50,000 - 74,999 & 55,314 - 78,289\\
    & 3& 50& 75,000 - 99,999 & 78,290 - 103,750\\
    & 4& 31& 100,000 + & 103,751\\ \hline
    \multirow{4}{*}{San Francisco} & 1& 33& $\leq$ 59,999 & $\leq$ 67,188\\
    & 2& 29& 60,000 - 99,999 & 67,189 - 95,300\\
    & 3& 35& 100,000 - 149,999 & 95,301 - 130976\\
    & 4& 38& 150,000 + & 130976 +\\ \hline
    \multirow{4}{*}{Seattle} & 1& 30& $\leq$ 49,999 & $\leq$ 54,306\\
    & 2& 23& 50,000 - 74,999 & 54,307 - 76,184\\
    & 3& 42& 75,000 - 99,999 & 76,185 - 99,722\\
    & 4& 37& 100,000 + & 99,723 +\\ \hline
    \multirow{4}{*}{Washington} & 1& 26& $\leq$ 74,999 & $\leq$ 76,734\\
    & 2& 63& 75,000 - 99,999 & 76,735 - 103,750\\
    & 3& 64& 100,000 - 149,999 & 103,751 - 134,955\\
    & 4& 58& 150,000 + & 134,956 +\\ \hline
    \end{tabular}
    \caption{ATUS income quartile breaks \cite{bls_atus_dd_cps_2017}.}
    \label{tab:atusincomebreaks}
\end{table}

\subsection*{}
\textbf{ATUS and average mobility data visitation.} Tables \ref{tab:atusvisitationstats} and \ref{tab:bostonMDvisitationstats} show basic information on ATUS and average mobility data visitation by POI category and income quartile. 

\begin{table}[H]
    \centering
    \begin{tabular}{|lccccc|} \hline
    \textbf{POI Category} &  \textbf{Inc. Quart.} & \textbf{Respondents} & \textbf{Respondents (\%)} & \textbf{Ave. Visit Length (min.)} & \textbf{Visits}\\ \hline
    \multirow{4}{*}{Food \& Coffee} & 1& 219& 21.2& 45.4& 345\\
    & 2& 119& 34.1& 46.9& 181\\ 
    & 3& 115& 30.8& 48.6& 192\\ 
    & 4& 279& 37.3& 47.8& 425\\ \hline
    \multirow{4}{*}{Grocery} & 1& 168& 16.3& 42.9& 178\\
    & 2& 61& 17.5& 42.3& 72\\
    & 3& 52& 13.9& 36.9& 63\\
    & 4& 152& 20.3& 40.8& 168\\ \hline
    \multirow{4}{*}{Gym} & 1& 31& 3.0& 52.6& 54\\
    & 2& 17& 4.87& 46.9& 31\\
    & 3& 23& 6.2& 46.0& 35\\
    & 4& 54& 7.2& 47.4& 90\\ \hline 
    \end{tabular}
    \caption{ATUS POI category visitation by income quartile displays significant variability in respondent time spent. For the 2017 ATUS: quartile 1 has 1,031 respondents, quartile 2 has 349 respondents, quartile 3 has 373 respondents, and quartile 4 has 749 respondents \cite{bls_2017}.}
    \label{tab:atusvisitationstats}
\end{table}

\begin{table}[H]
    \centering
    \begin{tabular}{|lccccc|} \hline
    \textbf{POI Category} &  \textbf{Inc. Quart.} & \textbf{Users} & \textbf{Users (\%)} & \textbf{Ave. Visit Length (min.)} & \textbf{Visits}\\ \hline
    \multirow{4}{*}{Food \& Coffee} & 1& 33,964& 83.9& 34.8& 275,395\\
    & 2& 33,036& 81.8& 35.6& 259,164\\ 
    & 3& 32,540& 81& 35.7& 243,985\\ 
    & 4& 31,837& 79.7& 35.6& 228,065\\ \hline
    \multirow{4}{*}{Grocery} & 1& 15,142& 37.7& 23& 42,331\\
    & 2& 13,380& 33.3& 23.4& 35,326\\
    & 3& 12,112& 30.8& 22.9& 30,777\\
    & 4& 11,311& 28.8& 22.6& 26,067\\ \hline
    \multirow{4}{*}{Gym} & 1& 7,180& 17.8& 41.3& 21,558\\
    & 2& 7,428& 18.3& 41.5& 22,382\\
    & 3& 7,669& 19& 41.2& 22,646\\
    & 4& 8,126& 20.1& 41& 23,143\\ \hline 
    \end{tabular}
    \caption{Mobility data POI category visitation by quartile, averaged across the 11 study cities, displays significant variability in user time spent. 
    The number of average respondents for mobility data is 40,354 for quartile 1, 40,160 for quartile 2, 39,950 for quartile 3, and 39,950 for quartile 4.}
    \label{tab:bostonMDvisitationstats}
\end{table}

\subsection*{}
\textbf{Comparison reveals city-differences in ATUS / MD Ratios.}
Figure \ref{fig:fig_total_time_curves} compares ATUS minutes per hour spent against mobility data for Boston and the resulting ATUS / mobility data ratios by income quartile. In the Supplementary Information, Figures \ref{fig:figs1}--\ref{fig:figs3} compare the ratios for the 11 study cities, one figure for each POI category.  

\begin{figure}[H]
    \centering
    \includegraphics[width=1 \linewidth]{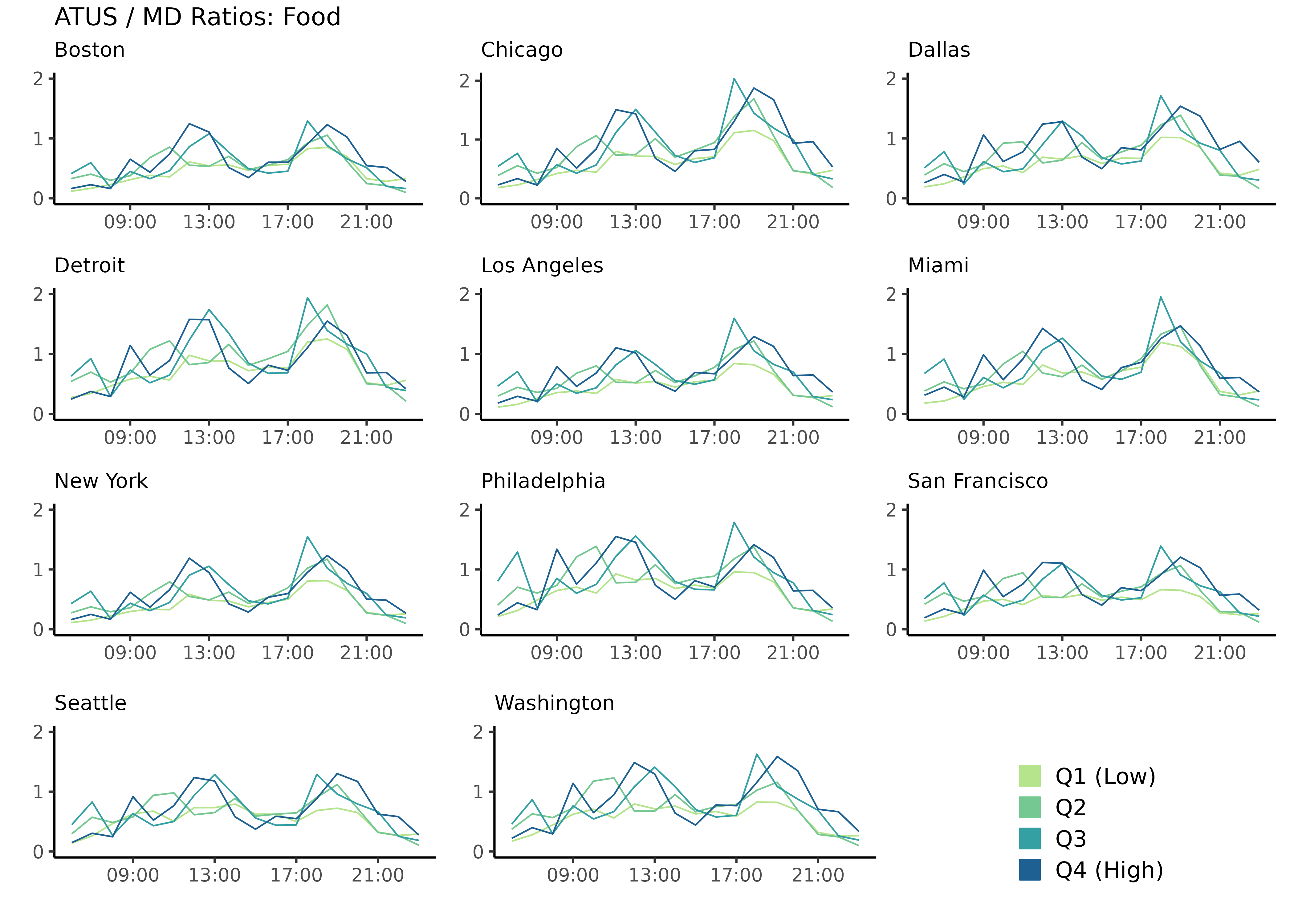}
    \caption{The ATUS / MD ratios by income quartile of the 11 cities from 6 a.m to 11 p.m. for Food and Coffee establishments.}
    \label{fig:figs1}
\end{figure}

\begin{figure}[H]
    \centering
    \includegraphics[width=1 \linewidth]{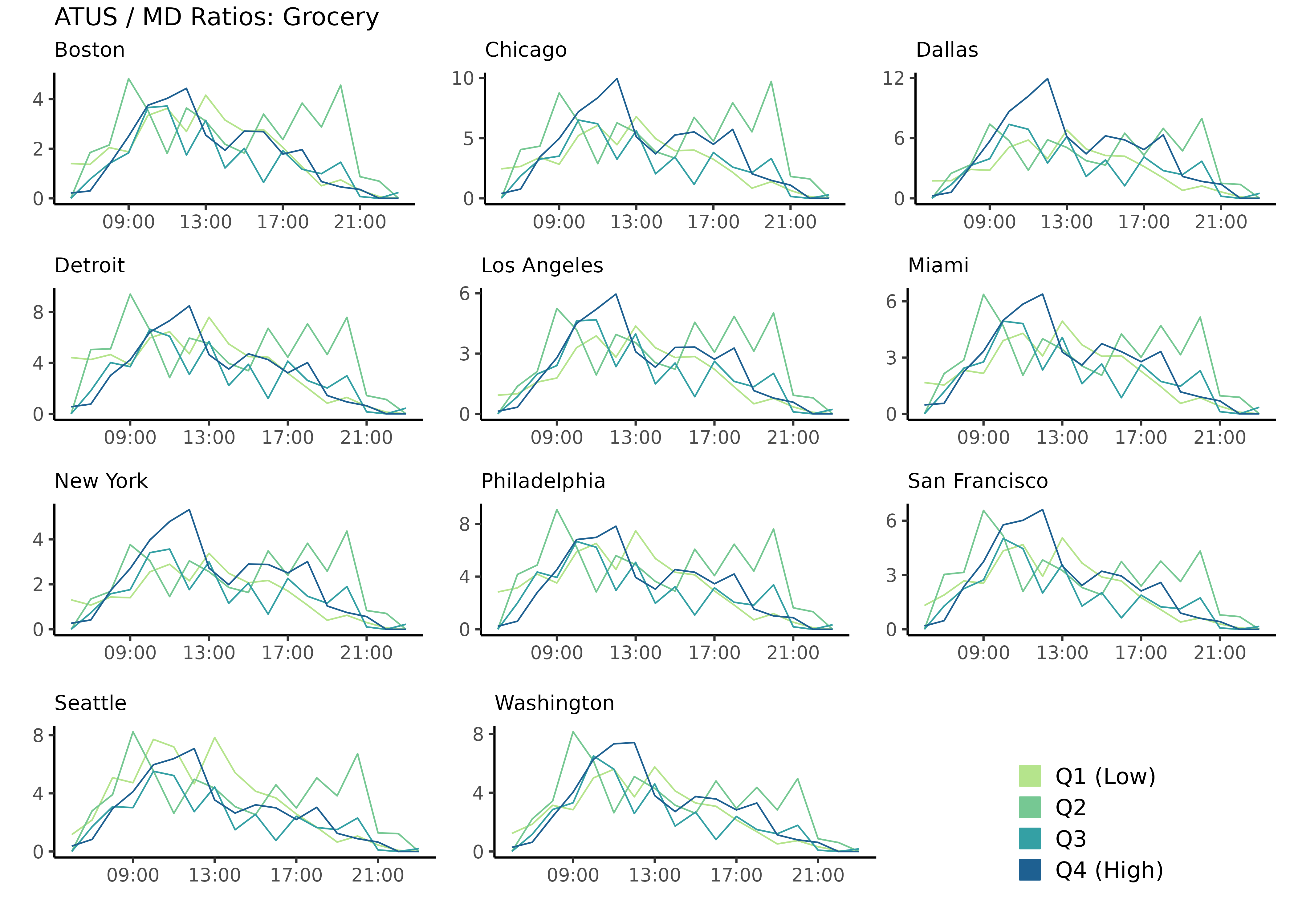}
    \caption{The ATUS / MD ratios by income quartile of the 11 cities from 6 a.m. to 11 p.m. for Grocery stores.}
    \label{fig:figs2}
\end{figure}

\begin{figure}[H]
    \centering
    \includegraphics[width=1 \linewidth]{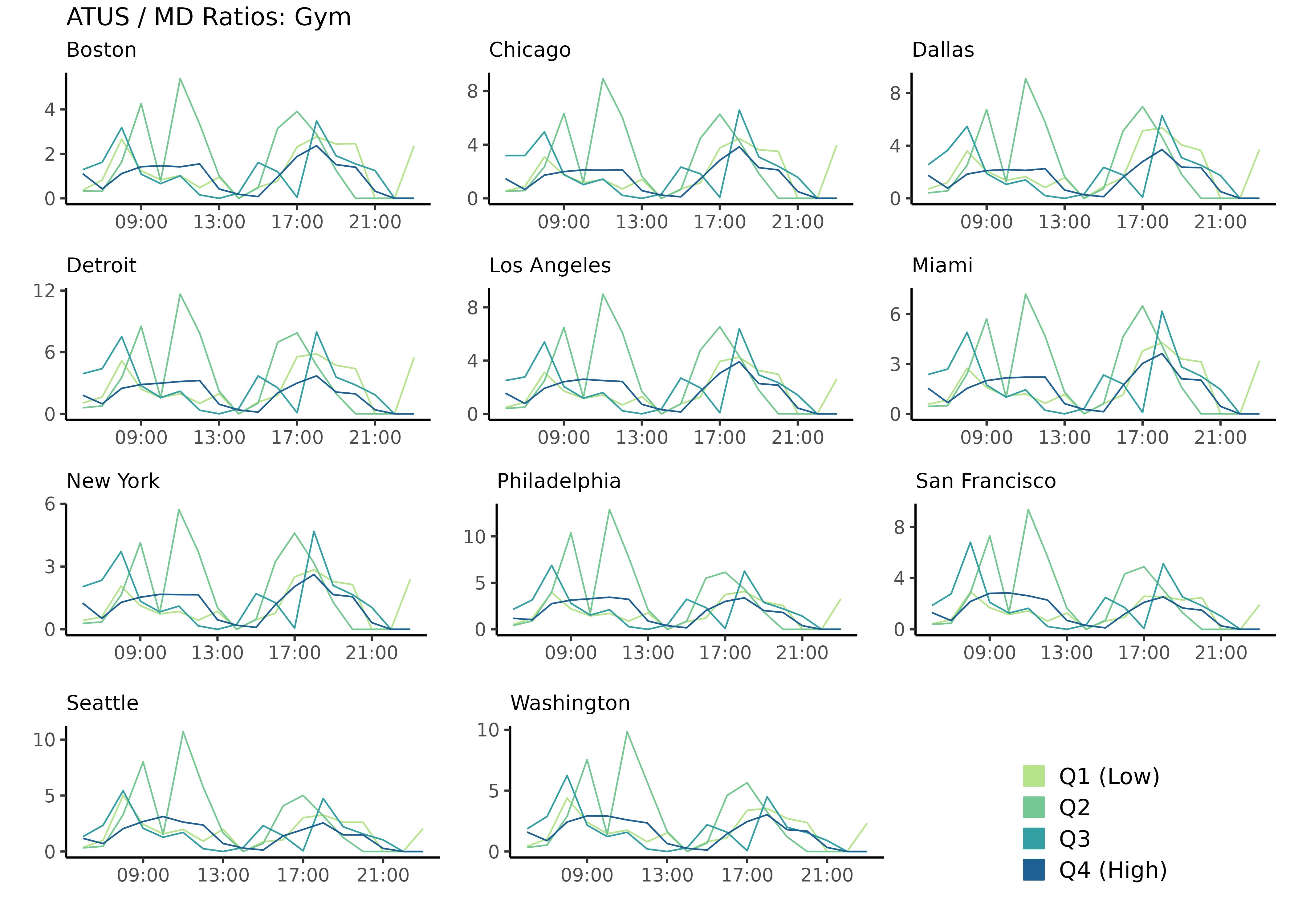}
    \caption{The ATUS / MD ratios by income quartile of the 11 cities from 6 a.m. to 11 p.m. for Gyms.}
    \label{fig:figs3}
\end{figure}

\subsection*{}
\textbf{Additional details on Figure 4.}
\begin{figure}[H]
    \centering
    \includegraphics[width=1 \linewidth]{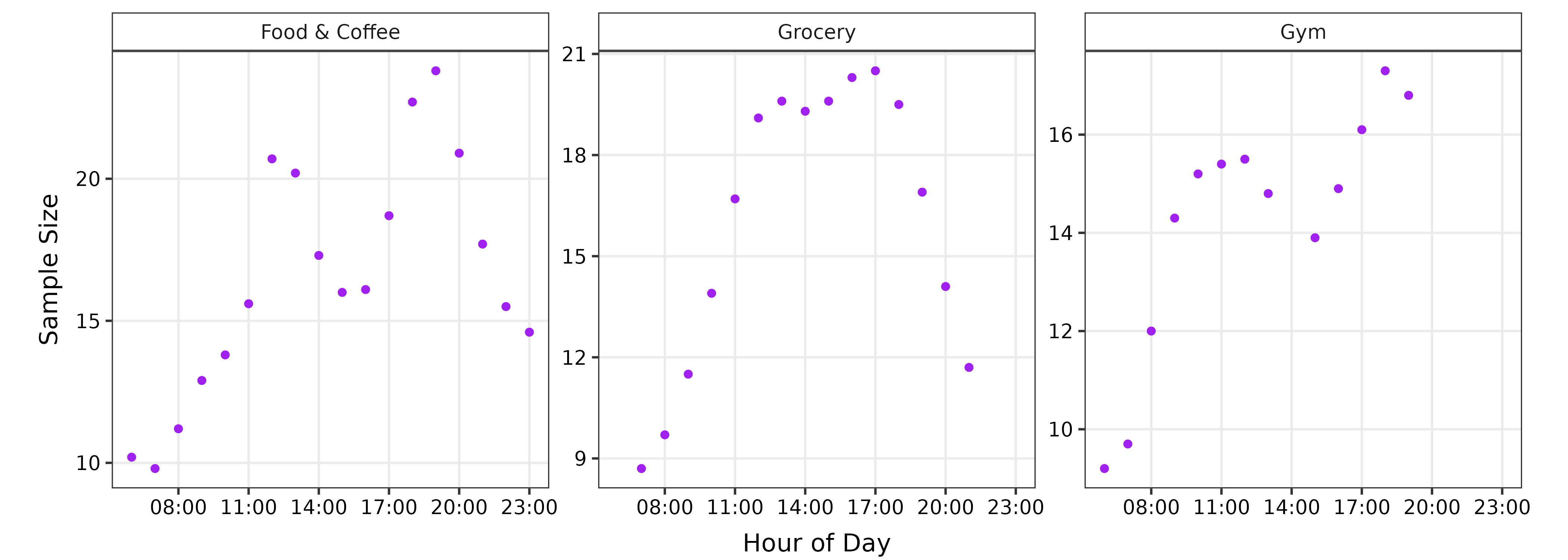}
    \caption{Diurnal variation of unique mobility data visitors per hour to each POI category averaged across cities.}
    \label{fig:addendumfig4}
\end{figure}

\subsection*{}
\textbf{Regression results by city.}
Table \ref{tab:mult_lin_reg} shows the multiple linear regression model results when the city is set as a fixed effect to examine the results without any influence from city-based differences. However, we were also interested in the regression results for the same variables for each city.

\begin{figure}[H]
    \centering
    \includegraphics[width=1 \linewidth]{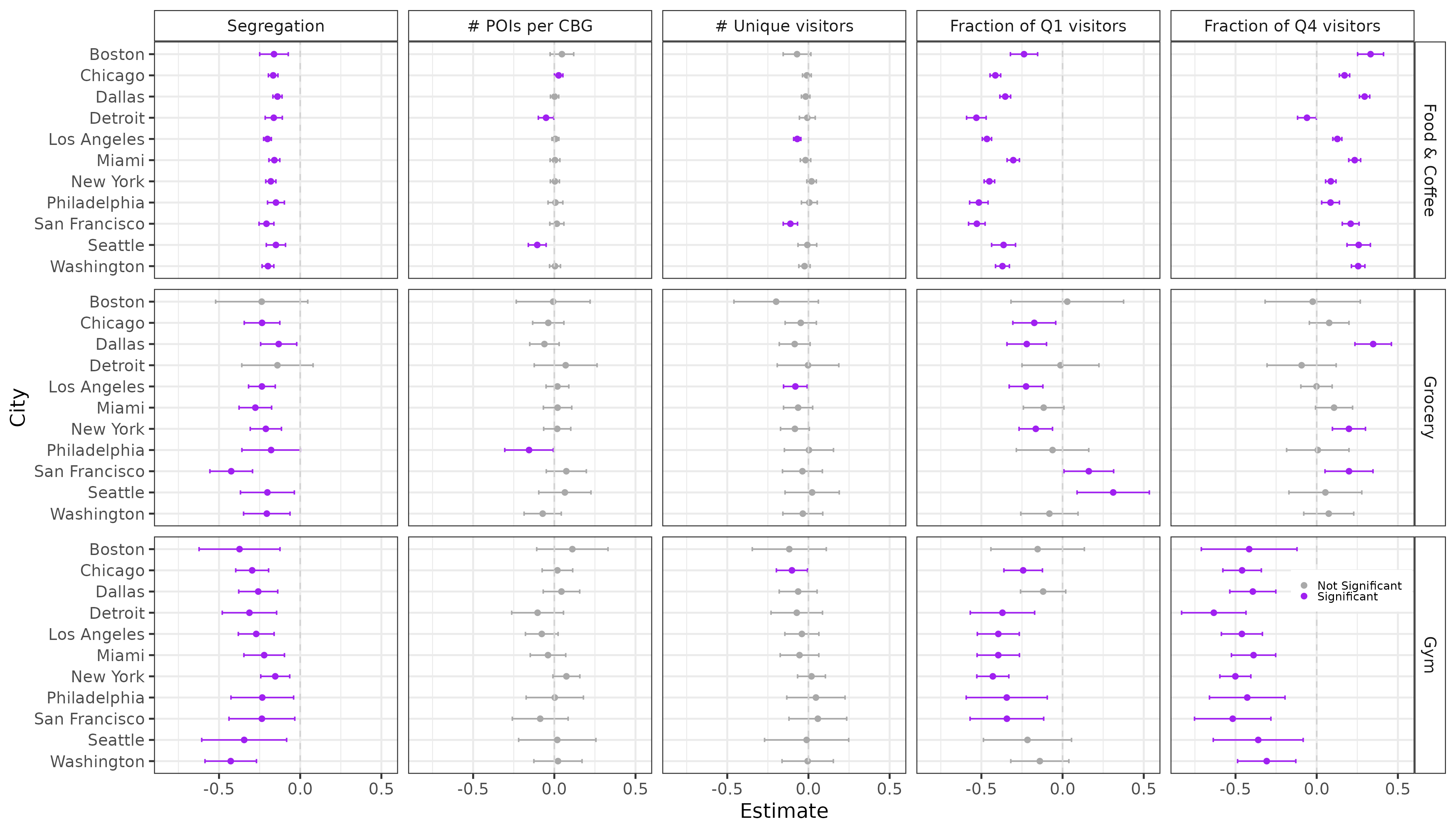}
    \caption{Multiple linear regression results by city.}
    \label{fig:regressionresults}
\end{figure}

\end{document}